\documentclass[11pt,journal,draftcls,onecolumn,peerreviewca]{IEEEtran}
\usepackage[T1]{fontenc}
\usepackage{algorithm}
\usepackage{algorithmic}
\usepackage{amsmath}
\usepackage{amsthm}
\usepackage{amssymb}
\usepackage{bbm}
\usepackage{color}
\usepackage{diagbox}
\usepackage{esint}
\usepackage{float}
\usepackage{graphicx}
\usepackage{mathrsfs}
\usepackage{microtype}
\usepackage{multicol}
\usepackage{stackrel}
\usepackage{stfloats}
\usepackage{stmaryrd}
\usepackage{subfigure}
\usepackage{verbatim}
\SetSymbolFont{stmry}{bold}{U}{stmry}{m}{n}

\begin{document}
\title{A Modulo Sampling Hardware Prototype and Reconstruction Algorithm Evaluation}
\author{Jiang Zhu, Junnan Ma, Zhenlong Liu, Fengzhong Qu, Zheng Zhu and Qi Zhang}
\maketitle
\begin{abstract}
  Analog-to-digital converters (ADCs) play a vital important role in any devices via manipulating analog signals in a digital manner. Given that the amplitude of the signal exceeds the dynamic range of the ADCs, clipping occurs and the quality of the digitized signal degrades significantly. In this paper, we design a joint modulo sampling hardware and processing prototype which improves the ADCs' dynamic range by folding the signal before sampling. Both the detailed design of the hardware and the recovery results of various state-of-the-art processing algorithms including our proposed unlimited sampling line spectral estimation (USLSE) algorithm are presented. Additionally, key issues that arise during implementation are also addressed. It is demonstrated that the USLSE algorithm successfully recovers the original signal with a frequency of $2.5$ kHz and an amplitude $10$ times the ADC's dynamic range, and the linear prediction (LP) algorithm successfully recovers the original signal with a frequency of $3.5$ kHz and an amplitude $10$ times the ADC's dynamic range.
\end{abstract}
\section{Introduction}
Analog-to-digital converters (ADCs) are omnipresent in digital devices, serving as the vital link between real-world analog signals and their digital counterparts. Typically, ADCs sample the analog signal uniformly and output the digital formats. Crucial parameters of the ADCs are the sampling frequency, the power consumption, and the dynamic range (DR). The DR refers to the ratio of the largest signal amplitude to the smallest detectable signal amplitude that the ADC can accurately resolve, which is ${\rm DR}=6B+1.73$ (dB) where $B$ is the bit-depth of the ADC. The sampling frequency determines how often the ADC samples the analog signal per unit of time. The power consumption depends on the bit-depth $B$, the sampling frequency and some other factors. 

In certain practical scenarios, the DR of the traditional ADC may prove insufficient as shown below:
\begin{itemize}
	\item Near-far field scenarios: Both weak and strong targets are present. If the DR of the ADC is tailored to match the amplitude of the strong target, signals from weaker targets might become indistinguishable from quantization noise, resulting in missed detection. Employing automatic gain control (AGC) to amplify the signal can prevent loss of weak signals; however, this approach risks saturating the ADC with the strong target's signal, leading to clipping at a level dictated by the power rails. This phenomenon also happens in wireless full-duplex (FD) transceivers where amplitude of the signal of interest (SoI) is much smaller than that of the strong self-interference (SI) \cite{OrdoñezfullduplexmoduloADC}. 
	\item The traditional ADC has the bottlenecks of power consumption and receiver saturation in Massive MIMO systems due to large radio frequency chains \cite{MassiveMIMOlambda}.
	\item The traditional biomedical imaging sensor \cite{sasagawa2015implantable} either has high effective peak signal-to-noise ratio (SNR) to detect a small intensity change, or has small dimensions and lightweight to be implanted in a mouse or a rat brain. 
\end{itemize}
These challenges have motivated researchers to investigate more efficient sampling architectures. One solution is the modulo sampling which folds the original signal before sampling. 
\subsection{Related Work}
The related modulo sampling works can be classified into three categories: Theory, algorithms and hardware design. Although we detail these separately, it is worth noting that these are complement to each other.

From the theoretical side of view, the tradeoff between the mean squared error (MSE) distortion and the required quantization rate, the sampling rate required for unique recovery of the bandlimited signals from modulo samples is investigated in depth. In \cite{ordentlich2018modulo}, a modulo-based architecture for ADC is proposed, where the converter is universal and agnostic of the statistics of the signal, the decoder utilizes the second order statistics (SOS) of the signal to unwrap the modulo folding.  In addition, identical modulo ADCs are applied componentwisely on each component of the random vectors, and the performance of the suboptimal integer forcing (IF) decoder is analyzed.  In \cite{bhandari2020unlimited},  a constructive recovery algorithm that is guaranteed to recover the underlying signal for oversampling rate $2\pi {\rm e}$ compared to the Nyquist rate is proposed. In \cite{bhandari2019identifiability}, it is shown that any sampling rate with modulo folding that is faster than the Nyquist sampling allows for injectivity over the class of finite energy bandlimited functions. Also, \cite{romanov2019above} independently and concurrently obtains the same identifiability conditions, and provides an explicit linear prediction (LP) based algorithm for unwrapping modulo folded signals. In \cite{prasanna2020identifiability}, necessary and sufficient conditions for unique recovery of sparse signals in the modulo compressed sensing setup are derived, and the minimum number of measurements to uniquely reconstruct every $s$-sparse signal from modulo measurements is $2s+1$. In \cite{ordentlich2016integer, APB2020}, it is shown that in the informed case where the covariance matrix ${\boldsymbol\Sigma}$ of the random Gaussian vectors is known, the error probability in reconstructing the original signal from the modulo signal depends on the ratio $\Delta/{\rm det}^{1/(2K)}({\boldsymbol\Sigma})$, where $K$ is the dimension of the signal. Later in \cite{weiss2022blind}, a blind version of the modulo architecture that does not need the SOS of the signal is proposed. Later, \cite{romanov2021blind} shows that the performance of the proposed blind recovery
algorithm closely follows that of the informed one, provided that the informed decoder recovers each sample with small error probability and the number of samples is large. In \cite{mulleti2024modulo},  based on number theory theorem, it is shown that the periodic bandlimited signal is identifiable from modulo measurements provided that the number of measurements is a prime number. Later in \cite{Qi2023ID}, an equivalent condition for identifying the signal from modulo measurements is proposed for any integer. The Cramér-Rao bounds (CRBs) are derived for both quantized and unquantized cases, considering scenarios with known and unknown folding counts, providing performance benchmarks \cite{cheng2023crb}.

From the algorithmic point of view, the time domain or the frequency domain properties of the signal is utilized to recover the original signal. In \cite{ordentlich2018modulo}, a linear prediction based approach is proposed, and it is shown to approach information theoretical limits, as captured by the rate-distortion function, in various settings. In \cite{bhandari2020unlimited}, an unlimited sampling algorithm (USAlg) is proposed which utilizes the fact that the higher order differences of the bandlimited samples are small enough in amplitude, so that the modulo of the higher order differences of modulo samples will be the same as the higher order differences of the bandlimited samples. Since then, USAlg has been extended to tackle many problems, such as the mixture of sinusoids, finite-rate-of-innovation signals, multidimensional signals, DOA estimation, computed tomography, graph signals
\cite{bhandari2018unlimited1,bhandari2018unlimited2,bouis2021multidimensional,fernandez2021doa,bhandari2021modulo,fernandez2022computational,beckmann2020hdr,bhandari2020hdr}, etc. However, due to using the higher-order differences, USAlg is sensitive to noise, especially when the noise is added to the original signal before modulo sampling. In \cite{zhang2024line}, the unlimited sampling line spectral estimation (USLSE) algorithm, which integrates the dynamic programming (DP) and the orthogonal matching pursuit (OMP) methods, is proposed to recover the original signal from noisy modulo samples, and its performance is demonstrated via substantial numerical simulations and real data acquired by the mmWave radar. \cite{ordentlich2016integer, APB2020}  propose an integer forcing decoder to unwrap the modulo samples, and extend to the blind setup later in \cite{romanov2021blind}. It is worth noting that the main idea of the integer forcing is that one tries to design a invertable integer matrix such that the dynamic range of the original high dynamic range samples after the linear transformation is as small as possible, such that the modulo of the linear transform of the modulo samples equals to the linear transform of the original samples, and one is hopefully to unwrap the original samples. In \cite{Qi2023onebit}, the authors propose a one bit aided modulo sampling to perform the DOA estimation in high dynamic range environments, and show that the weak signal can be detected reliably. With the knowledge of the number of folds, \cite{bhandari2021unlimited} proposed a frequency domain algorithm for recovering the periodic bandlimited signals through solving a spectral estimation problem with Prony's method, and this method can be used to handle the non-ideal foldings in practical hardware. \cite{azar2022residual,azar2022robust} uses the projected gradient descent method to estimate the residual in the Fourier domain via oversampling.

From the hardware point of view, a phase-domain implementation of modulo ADC via ring oscillators is developed \cite{ordentlich2018modulo}, where the input voltage is first converted into phase and then is quantized, which is very natural in implementing the modulo operation. In \cite{bhandari2020unlimited, bhandari2021unlimited, florescu2022surprising, florescu2022time}, the modulo hardware prototypes where the modulo operation is implemented prior to sampling, and the hardware limitations such as the hysteresis effects and the performance of the algorithms are focused, rather on the detailed hardware implementation. It has been demonstrated that the modulo hardware can fold low-frequency signals (below $300$ Hz) with amplitudes up to $10$ times greater than the ADC's dynamic range. Later in \cite{Krishnacircuit2019}, the unlimited dynamic range ADC (UDR-ADC) consisting of the sample-and-hold circuit, modulo circuit and quantizer is designed. Two bits which provide the side information are employed to encode the three reset possibilities (positive reset, negative reset, or no reset). The UDR-ADC is similar to the conventional ADC in that the modulo sampling and quantization is used after the sampling and hold circuit, which results in a slower ADC due to a larger holding time. It is shown that the resulting hardware circuit is able to fold signals on the order of Hz with the signal's amplitude being slightly larger than three times the ADC's dynamic range. In \cite{mulleti2023hardware}, a modulo hardware prototype for sampling of signals up to 10 kHz is presented, and the bandlimited and FRI signals recovered by the beyond bandwidth residual reconstruction ($B^2R^2$) algorithm \cite{azar2022residual} is demonstrated. However, all the modulo hardware prototypes use the oscilloscope to acquire the sampled data, instead of performing sampling on the modulo samples.

\subsection{Main Contributions}
In this paper, we present a modulo sampling prototype to perform modulo sampling, which uses the ADC to sample the signal instead of acquiring the data through oscilloscope. The full design details of the modulo sampling prototype are provided. In addition, we also evaluate various state-of-the-art algorithms including our proposed USLSE \cite{zhang2024line}, USAlg \cite{bhandari2020unlimited}, and LP \cite{ordentlich2018modulo} algorithms that recover the signal from modulo samples, where the data is acquired by the designed modulo sampling prototype. It is shown that both USLSE and LP are robust in estimating the signal from the modulo samples, which validates the effectiveness of the joint design of the modulo sampling hardware prototype and the reconstruction algorithm.

\section{Modulo Sampling and Benchmark Algorithms}
In this section, we consider modulo sampling for signals whose amplitude is higher than the dynamic range of the ADC. The number of folds for the signals can be very large, but the recovery designed for unfolding the signal become difficult when the number of folds is large, especially in the presence of noise before modulo sampling. 

Consider a bandlimited signal $s(t)$ such that its bandwidth is $B$, i.e., the support of the Fourier transform $S(f)$ of $s(t)$ is $[-B,B]$. According to the Nyquist criterion, the signal can be perfectly reconstructed from the Nyquist rate $f_{\rm Nyq}=2B$ samples per second, provided that the DR of the ADC is larger than the DR of the signal. Given that the DR of the signal exceeds that of the ADC, the ADC will be saturated and the signal will be clipped, and distortion occurs. To avoid clipping, the modulo sampling is proposed where a modulo operator is applied before sampling. In this way, the modulo operation can be mathematically described as  
\begin{align}\label{fold_obs}
	y(t)\triangleq \mathscr{M}_{\lambda}\left(g(t)\right)\triangleq \mathscr{M}_{\lambda}\left(s(t)+w(t)\right),
\end{align}
where $w(t)$ is the additive noise, $g(t)$ is the noisy original signal, $\mathscr{M}_{\lambda}\left(\cdot\right)$ is the centered modulo mapping defined as
\begin{align}
	\mathscr{M}_{\lambda}\left(x\right)\triangleq x - 2\lambda\left\lfloor\frac{x}{2\lambda} + \frac{1}{2}\right\rfloor,
\end{align}
$\lfloor \cdot \rfloor$ denotes the floor operator, and $\lambda$ denotes the threshold of ADC. Note that the original noisy signal $g(t)$ can be decomposed as the sum of modulo signal and a simple function \cite{bhandari2020unlimited}, i.e.,
\begin{align}\label{fold_obs_decom}
	g(t) = s(t) + w(t) = y(t) + 2\lambda\epsilon(t),
\end{align}
where the range of the simple function $\epsilon(t)$
is $\mathbb{Z}$ which is the set of integers. Sampling $y(t)$ with frequency $f_s$ yields
\begin{align}\label{samplingmeas}
y[n]&=y(t)|_{t=nT_s} = \mathscr{M}_{\lambda}\left(g(t)|_{t=nT_s}\right) \notag\\
    &= \mathscr{M}_{\lambda}\left(g[n]\right) = g[n] - 2\lambda \epsilon[n],
\end{align}
where $n=0,1,\cdots,N-1$ and $N$ is the number of samples. In this paper, the signal is oversampled, and the oversampling factor denoted by $\gamma$ is defined as
\begin{align}
   \gamma = \frac{f_s}{f_{\rm Nyq}} = \frac{f_s}{2B}.
\end{align}

Let ${\mathbf g} \triangleq [g[0],\cdots,g[N-1]]^{\rm T}$, ${\mathbf y}\triangleq [y[0],\cdots,y[N-1]]^{\rm T}$, and ${\boldsymbol{\epsilon}}\triangleq [\epsilon[0],\cdots,\epsilon[N-1]]^{\rm T}$, model (\ref{samplingmeas}) can be reformulated as 
\begin{align}\label{vectormodel}
	\mathbf{y} = \mathbf{g} - 2\lambda\boldsymbol{\epsilon},
\end{align}
and the primary goal in modulo sampling is to recover the unfolded samples $\mathbf{g}$ from modulo samples ${\mathbf y}$. In this paper, we design and implement a hardware prototype of modulo sampling, referred to as a modulo ADC. Several benchmark algorithms are employed to recover the unfolded samples $\mathbf{g}$ from the modulo samples $\mathbf{y}$ obtained from the modulo ADC. Below, we briefly introduce these benchmark algorithms.
\subsection{USAlg}\label{USAlgSec}
For the USAlg algorithm \cite{bhandari2020unlimited}, the higher order differences operator is first applied to the modulo samples $\mathbf{y}$, i.e.,
\begin{align}
  \mathbf{y}^{(D_O)} = \mathbf{J}^{(N-D_O)}\cdots \mathbf{J}^{(N-2)}\mathbf{J}^{(N-1)}\mathbf{y},
\end{align}
where $\mathbf{J}^{(N-i)} = \mathbf{I}^{(N-i+1)}_{2:N-i+1,:} - \mathbf{I}^{(N-i+1)}_{1:N-i,:}$ is the first-order difference operator of a sequence with length $N-i+1$ for $i=1,2,\cdots, D_O$, and $D_O$ is the order of difference. Here, $\mathbf{I}^{(N-i+1)}$ is the identity matrix of dimension $N-i+1$. According to the measurement model (\ref{vectormodel}), we have 
\begin{align}\label{HoDvectormodel}
    \mathbf{y}^{(D_O)} = \mathbf{g}^{(D_O)}-2\lambda\boldsymbol{\epsilon}^{(D_O)},
\end{align}
where $\mathbf{g}^{(D_O)}$ and $\boldsymbol{\epsilon}^{(D_O)}$ are the sequences obtained by applying the $D_O$th-order difference operator to $\mathbf{g}$ and $\boldsymbol{\epsilon}$. In addition, based on Lemma $2$ in \cite{bhandari2020unlimited}, in the noiseless case, when the oversampling factor $\gamma$ is greater than $2\pi e$, each element of $\mathbf{g}^{(D_O)}$ will be constrained within the dynamic range of the ADC, i.e., $\|\mathbf{g}^{(D_O)}\|_{\infty} \leq \lambda$, provided that
\begin{align}
    D_O \geqslant \left\lceil \frac{\log \lambda - \log \beta_g}{\log \left(\frac{\pi e}{\gamma}\right)} \right\rceil,
\end{align}
where $\lceil\cdot\rceil$ is the ceiling operator and $\beta_g$ is a known constant that satisfies
\begin{align}
    \beta_g \in 2 \lambda \mathbb{Z} \quad \text{and} \quad \|\mathbf{g}\|_{\infty} \leqslant \beta_g. 
\end{align}
Thus, by applying the modulo operator to (\ref{HoDvectormodel}), we have 
\begin{align}
    \mathscr{M}_{\lambda}(\mathbf{y}^{(D_O)})= \mathscr{M}_{\lambda}(\mathbf{g}^{(D_O)}) = \mathbf{g}^{(D_O)}
\end{align}
since $\mathscr{M}_{\lambda}(2\lambda\boldsymbol{\epsilon}^{(D_O)}) = \mathbf{0}$ and $\|\mathbf{g}^{(D_O)}\|_{\infty} \leq \lambda$. Replacing $\mathbf{g}^{(D_O)}$ with $\mathscr{M}_{\lambda}(\mathbf{y}^{(D_O)})$ in (\ref{HoDvectormodel}), we have 
\begin{align}
    2\lambda\boldsymbol{\epsilon}^{(D_O)} = \mathscr{M}_{\lambda}(\mathbf{y}^{(D_O)}) - \mathbf{y}^{(D_O)}.\notag
\end{align}
Furthermore, the simple function $\boldsymbol{\epsilon}$, up to an additive constant, is estimated from $\boldsymbol{\epsilon}^{(D_O)} = (\mathscr{M}_{\lambda}(\mathbf{y}^{(D_O)}) - \mathbf{y}^{(D_O)}) / 2\lambda$ by applying the anti-difference operator. Finally, given the estimate of the simple function $\boldsymbol{\epsilon}$, the original signal can be recovered. It should be noted that the USAlg algorithm is sensitive to noise, as the higher-order difference operator amplifies noise, potentially violating the inequality $\|\mathbf{g}^{(D_O)}\|_{\infty} \leq \lambda$, which adversely affects estimation performance.
\subsection{LP}
In the LP method \cite{ordentlich2018modulo}, the auto-covariance function is assumed to be known, and the $P$-tap prediction filter coefficients $\{h_p\}_{p=1}^{P}$ are computed based on it. Additionally, the initial $P$ unfolded samples $\{{g}_p\}_{p=1}^P$ are also needed to be given. The $n$th unfolded sample for $n = P+1, P+2, \dots, N$ is then recursively estimated. In detail, for the $n$th unfolded sample where $n = P+1, P+2, \dots, N$, it is first estimated as
\begin{align}
    \hat{g}_{n} = \sum_{i=1}^{P} h_i \hat{g}_{n - i}
\end{align}
where $\hat{g}_p = {g}_p$ for $p=1, 2, \cdots, P$. This estimate is then refined using the modulo sample $y_n$ as
\begin{align}
    \hat{g}_{n} :=  \hat{g}_{n} + \mathscr{M}_{\lambda}(y_{n} - \hat{g}_{n}), n = P+1, P+2, \dots, N.
\end{align}

\subsection{USLSE}
As shown in \cite{zhang2024line}, the USLSE algorithm is designed for LSE via modulo sampling under an oversampling setup \footnote{The USLSE code is available at https://github.com/RiverZhu/USLSEcode.}. The algorithm consists of two stages: In the first stage, it combines a DP method with the OMP to recover the original signal from the modulo samples; in the second stage, a state-of-the-art LSE algorithm is applied to estimate the parameters. In this paper, we use only the first stage of the USLSE algorithm. Although USLSE was originally applied to complex signals \cite{zhang2024line}, it can also be directly used for real signals. 

The first stage of the USLSE algorithm is briefly outlined below. First, the first-order difference operator followed by a discrete Fourier transform (DFT) is applied to the measurement model (\ref{vectormodel}), resulting in:
\begin{align}\label{meamodeldiffDFT}
	\widetilde{\mathbf{y}} = \widetilde{\mathbf{g}} - 2\lambda\mathbf{F}^{(N-1)}\underline{\boldsymbol{\epsilon}},
\end{align}
where $\widetilde{\mathbf{y}} \triangleq \mathbf{F}^{(N-1)}\mathbf{J}^{(N-1)}\mathbf{y}$, $\widetilde{\mathbf{g}} \triangleq \mathbf{F}^{(N-1)}\mathbf{J}^{(N-1)}\mathbf{g}$, and $\underline{\boldsymbol{\epsilon}} \triangleq \mathbf{J}^{(N-1)}\boldsymbol{\epsilon}$. Here, $\mathbf{J}^{(N-1)}$ is the first-order difference operator of a sequence with length $N$ which has been defined in Sec. \ref{USAlgSec} and $\mathbf{F}^{(N-1)}$ is the DFT matrix of dimension $N-1$. For simplicity, we use $\mathbf{F}$ instead of $\mathbf{F}^{(N-1)}$ in the following. Because of oversampling, the leakage of the bandlimited signal into the spectrum between the maximum frequency in the signal and the half of the sampling frequency can be controlled, the following approximate equation
\begin{align}
    2\lambda \mathbf{F}_{\mathcal{S}}\underline{\boldsymbol{\epsilon}} \approx -{\widetilde{\mathbf{y}}}_{\mathcal{S}}
\end{align}
is obtained, where $\mathcal{S}$ is a subset of ${1,2,\dots,N-1}$, and $\mathbf{F}_{\mathcal{S}}$ is a submatrix of $\mathbf{F}$ formed by selecting the rows of $\mathbf{F}$ corresponding to the indices in $\mathcal{S}$. Furthermore, leveraging the benefits of oversampling, where the amplitude of each element in $\underline{\boldsymbol{\epsilon}}$ is bounded, an optimization problem with respect to $\underline{\boldsymbol{\epsilon}}$ is formulated using the least squares method:
\begin{subequations}\label{opt_diff_sub}
	\begin{align}
		&\underset{\underline{\boldsymbol{\epsilon}}}{\operatorname{minimize}}~ \|{\widetilde{\mathbf{y}}}_{\mathcal{S}} + 2\lambda \mathbf{F}_{\mathcal{S}} \underline{\boldsymbol{\epsilon}}\|_2^2, \\
		&{\rm s.t.}~ \underline{\epsilon}_n \in \mathcal{V},~ n = 1, \ldots, N-1,
	\end{align}
\end{subequations}
where $\mathcal{V} = \{m \mid -V \leq m \leq V,~ m \in \mathbb{Z}\}$ is a bounded lattice. Additionally, $\mathbf{F}^{\rm H}_{\mathcal{S}}\mathbf{F}_{\mathcal{S}}$ is approximated as a $p$th-order diagonal matrix $\mathbf{A}$, where $\mathbf{A}_{i,j} = 4\lambda^2\mathbf{F}^{\rm H}_{i}\mathbf{F}_{j}$ if $|i-j| \leq p$, and $\mathbf{A}_{i,j} = 0$ otherwise, for any $1 \leq i, j \leq N-1$. Consequently, instead of solving optimization problem \eqref{opt_diff_sub}, an approximate optimization problem is formulated as:
\begin{subequations}\label{opt_diff_sub_appro}
	\begin{align}
		&\underset{\underline{\boldsymbol{\epsilon}}}{\operatorname{minimize}}~ \underline{\boldsymbol{\epsilon}}^{\rm H} \mathbf{A} \underline{\boldsymbol{\epsilon}} + 2 \Re\{\mathbf{b}^{\rm H} \underline{\boldsymbol{\epsilon}}\}, \\
		&{\rm s.t.}~ \underline{\epsilon}_n \in \mathcal{V},~ n = 1, \ldots, N-1,
	\end{align}
\end{subequations}
where $\mathbf{b} = 2\lambda \mathbf{F}^{\rm H}_{\mathcal{S}}\widetilde{\mathbf{y}}_{\mathcal{S}}$. The objective function in optimization problem (\ref{opt_diff_sub_appro}) exhibits a $p$th-order Markov property because $\epsilon_n$ is only related to variables $\epsilon_i$, where the index $i$ ranges from $\max\{n-p,1\}$ to $\min\{n+p,N-1\}$. Therefore, the global optimum for this problem can be efficiently obtained using a DP method. The solution of this problem serves as the initial estimate, which is further refined using the OMP method by directly solving the original optimization problem \eqref{opt_diff_sub}. In the first stage of the USLSE algorithm, several iterations of DP and OMP are performed. Finally, the estimate of $\boldsymbol{\epsilon}$, up to an additive constant, is obtained using the anti-difference operator, and the original signal $\mathbf{g}$ is then recovered based on the measurement model (\ref{vectormodel}).

\section{Hardware Prototype Design}\label{HardProtoDe}
\begin{figure*}[!ht]
	\centering
	\includegraphics[width=\textwidth]{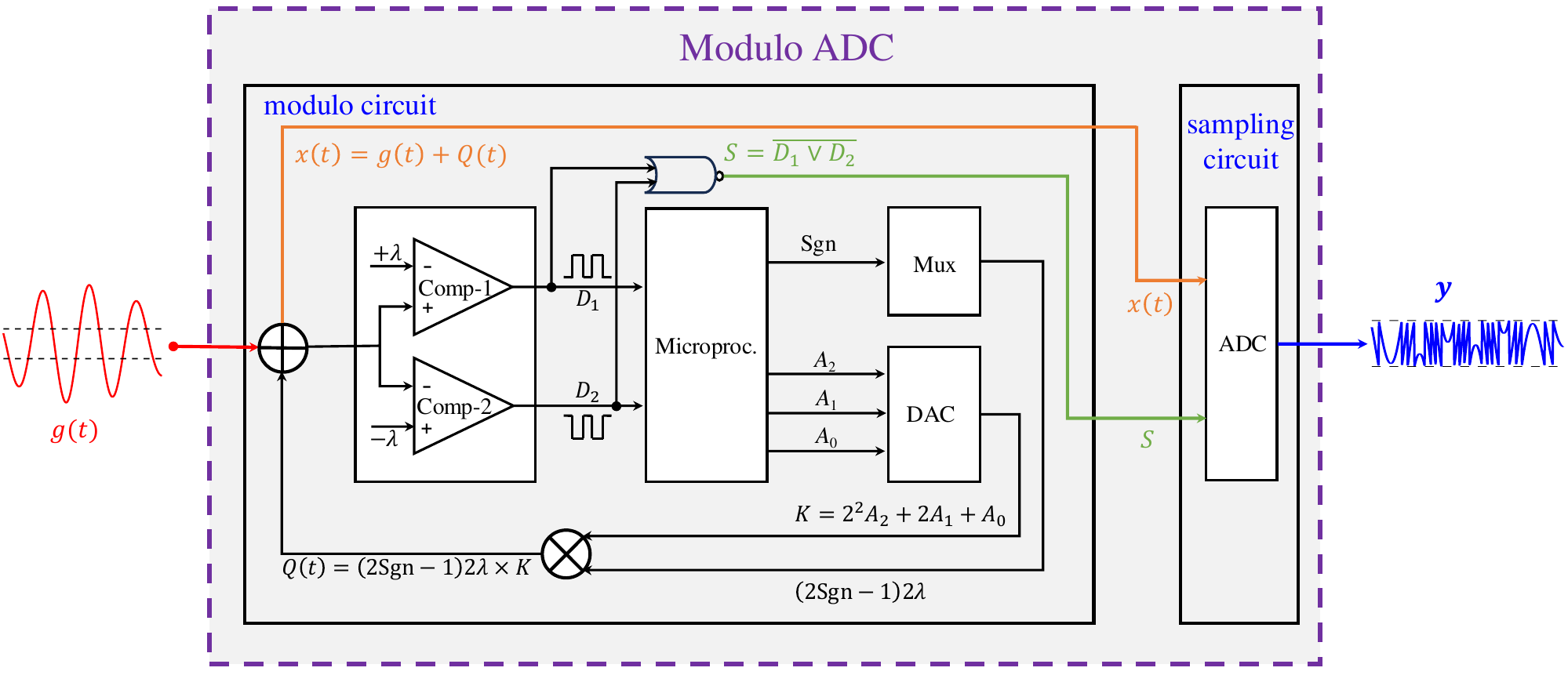}
	\caption{The architecture of the modulo ADC consists of two parts: the modulo circuit and the sampling circuit. In the modulo circuit, the input is the original high dynamic range signal $g(t)$, and the outputs are the modulo signal $x(t)$ and a control bit $S$ for sampling. In the sampling circuit, the inputs are the modulo signal $x(t)$ and the control bit $S$, while the output is the modulo samples $\mathbf{y}$.}
	\label{modADCarchitecture}
\end{figure*}

The modulo ADC hardware prototype utilizes the mixed analog and digital feedback circuit to fold the original high dynamic range signal into the low dynamic range signal, and then samples it to avoid saturation, see Fig. \ref{modADCarchitecture} for a schematic of the modulo ADC. 

First, we consider the modulo circuit. The input to the modulo circuit is the original high dynamic range signal $g(t)$, and the outputs are the modulo signal $x(t)$ and a control bit $S$ for sampling. The modulo signal $x(t)$ is obtained by adding $Q(t)$ to $g(t)$, where $Q(t)$ is a simple function constrained to take values in $2\lambda\mathbb{Z}$. In detail, $Q(t) = (2{\rm Sgn} - 1)2\lambda\times K$, where $(2{\rm Sgn} - 1)2\lambda$ and $K$ are the outputs of the multiplexer (Mux) and the digital-to-analog converter (DAC), respectively. ${\rm Sgn}$, which determines the sign of $Q(t)$, is a binary signal output from the microprocessor and serves as the input to the Mux. $K = 2^2A_2 + 2A_1 + A_0$ represents the number of folding counts, where $A_2$, $A_1$, and $A_0$ are binary signals output from the microprocessor and serve as inputs to the DAC. The maximum value of $K$ is $7$ when $A_2$, $A_1$, and $A_0$ are all equal to $1$. The control bit $S$ is the output of a NOR gate applied to $D_1$ and $D_2$, i.e., $S = \overline{D_1  \vee D_2}$, where $D_1$ is the output of comparator $1$ (Comp-$1$) and $D_2$ is the output of comparator $2$ (Comp-$2$). To better understand the working principles of the modulo circuit, let us consider a specific time instant $t_0$. At this instant, the multiplier outputs $Q(t_0)$, and the modulo signal output by the adder is $x(t_0) = g(t_0) + Q(t_0)$. When compared to $\pm\lambda$, it has three possible outcomes:
\begin{itemize}
	\item If $x(t_0)$ is larger than $\lambda$, i.e., $x(t_0)>\lambda$, then Comp-$1$ generates a high level voltage and Comp-$2$ generates a low level voltage, i.e., $D_1 = 1$ and $D_2 = 0$. 
	\item If $x(t_0)$ is smaller than $-\lambda$, i.e., $x(t_0)<-\lambda$, then Comp-$1$ generates a low level voltage and Comp-$2$ generates a high level voltage, i.e., $D_1 = 0$ and $D_2 = 1$. 
	\item If $x(t_0)$ is within $[-\lambda,\lambda)$, i.e., $|x(t_0)|<\lambda$, then both Comp-$1$ and Comp-$2$ generates low level voltages, i.e., $D_1 = D_2 = 0$. 
\end{itemize}
In the first case, when Comp-$1$ triggers a positive value, the microprocessor increments the folding number by $1$ in its binary representation. Through the DAC, the folding number in decimal form, $K$, is also incremented by $1$. The multiplier then updates the function $Q(t)$ by multiplying $K$ with $(2{\rm Sgn} - 1)2\lambda$. The control bit $S$ is set to $\overline{1 \vee 0} = 0$. In the second case, when Comp-$2$ triggers a positive value, the microprocessor decrements the folding number by $1$ in its binary representation, and the output of the DAC, $K$, is also decremented by $1$. $Q(t)$ is similarly updated by the multiplier. The control bit $S$ is set to $\overline{0 \vee 1} = 0$. In the third case, when both Comp-$1$ and Comp-$2$ generate low-level voltages, the function $Q(t)$ remains unchanged. The control bit $S$ is set to $\overline{0 \vee 0} = 1$. For the sampling circuit, the inputs are the modulo signal $x(t)$ and the control bit $S$, and the output is the modulo samples $\mathbf{y}$. Only when $S = 1$, i.e., when both Comp-$1$ and Comp-$2$ generate low-level voltages (indicating that $x(t)$ is within the range $[-\lambda,\lambda)$), does the ADC sample the modulo signal $x(t)$.

In the modulo ADC, the modulo signal $x(t)$ and the simple function $Q(t)$ are not equal to the ideal one in (\ref{fold_obs_decom}), i.e., $y(t)$ and $-2\lambda\epsilon(t)$\footnote{Note that in equation (\ref{fold_obs_decom}), the relationship is given by $y(t) = g(t) - 2\lambda\epsilon(t)$, whereas in the context of a modulo ADC, the expression becomes $x(t) = g(t) + Q(t)$.} due to the time delay in the feedback loop, which has also been discussed in \cite{mulleti2023hardware}. Let $\Delta T$ be the delayed time, we have $-2\lambda\epsilon(t-\Delta T) = Q(t)$. Thus, we have 
\begin{align}
    x(t) = g(t) + Q(t) = g(t) -2\lambda\epsilon(t-\Delta T).
\end{align}
Note that $2\lambda\epsilon(t-\Delta T)$ is the quantity on the lattice $2\lambda\mathbb{Z}$ to make sure that $g(t-\Delta T) -2\lambda\epsilon(t-\Delta T)$ is within the dynamic range $[-\lambda, \lambda]$. However, $x(t)$ may not in the dynamic range $[-\lambda, \lambda]$ with $\lambda=1$, as shown in the ensuing real data acquired by the prototype. 

\subsection{Design of Two Comparators}
\begin{figure}[!ht]
	\centering
	\includegraphics[width=0.48\textwidth]{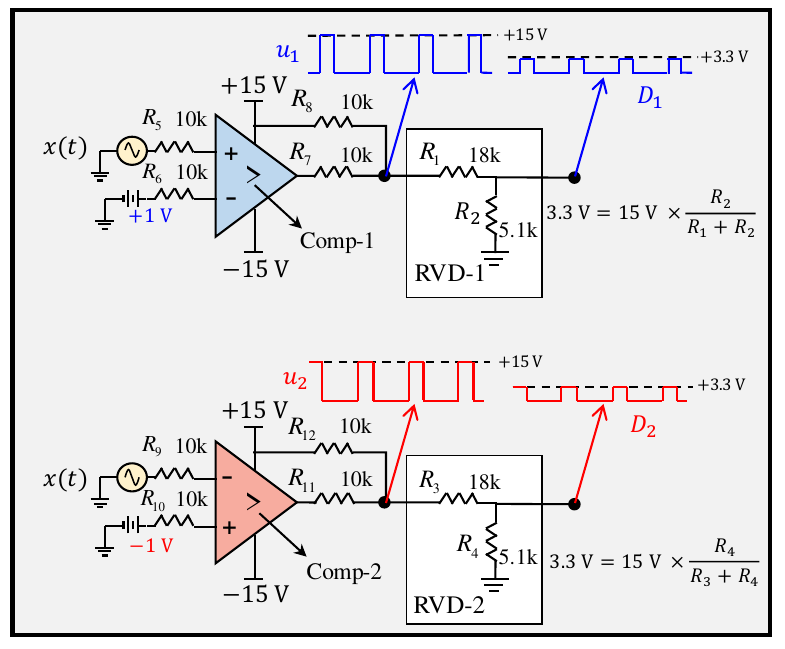}
	\caption{The detailed design of two comparators where $\lambda =1$. Comp-$1$: The $\lambda$ crossing detector where the input of the positive and negative side are $x(t)$ and $\lambda$, respectively; Comp-$2$: Comparing input voltages of opposite polarity where the input of the positive and negative side are $-\lambda$ and $x(t)$, respectively.}
	\label{Twocomparators}
\end{figure}

To determine whether the input voltage is greater than $\lambda$, within $[-\lambda,\lambda]$ or smaller than $-\lambda$, two comparators are needed. The first is a basic comparator, where the input $x(t)$ is compared with the reference voltage $\lambda$. The second compares input voltages of opposite polarity where the input of the positive and negative side are $-\lambda$ and $x(t)$, respectively. For the two comparators, their outputs are either $0$ or $V_{\rm CC}$. In our design, $V_{\rm CC}=15$ V. Note that the input voltage of the pins of the microprocessor ranges from $0$ V to $3.3$ V. Therefore, a  resistive voltage divider circuit is needed, which lowers the original high voltage $V_{\rm CC}=15$ V into $3.3$ V. The detailed design of two comparators are shown in Fig. \ref{Twocomparators} and $\lambda$ is set as $1$.

\subsection{Microprocessor}\label{micropro}
Here an STM32F103C8T6 is adopted to count the number of folds $K$, and decompose $K$ with an unsigned $3$-ary bits $A_2A_1A_0$ and the sign of $K$. It is worth noting that the fold number in the current instant is initialized and refined to update the fold number in the next instant. 

The second function of the microprocessor is to communicate with the host computer to transmit the samples of the folded signal. Here we use RS485 communication to transmit the samples collected by the ADC to the host computer, which requires the fewest number of pins, offers the highest resistance to interference, and has the fastest communication rate. 
\subsection{DAC Circuit}
\begin{figure*}[!ht]
	\centering
	\includegraphics[width=\textwidth]{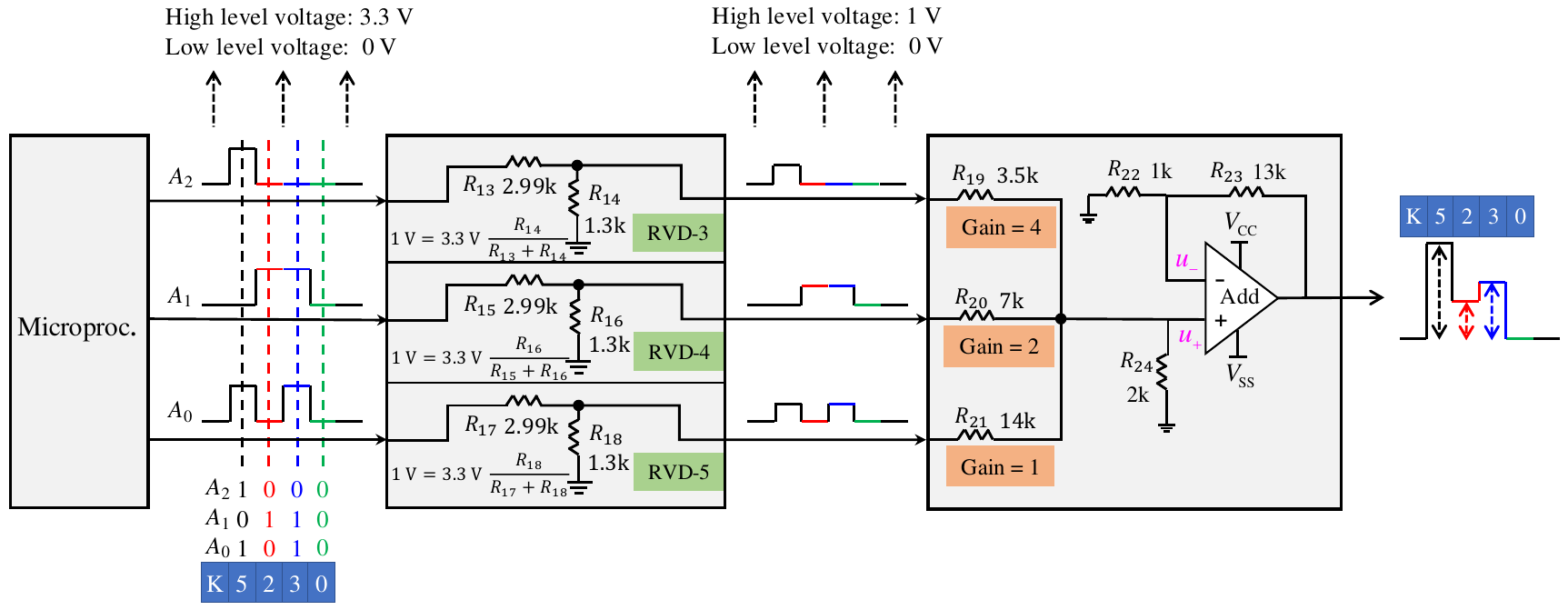}
	\caption{The detailed design of the DAC circuit. The inputs are three binary signals output
from the microprocessor, i.e., $A_2$, $A_1$ and $A_0$. The output is the number of folding times in decimal form, i.e, $K = 2^2A_2 + 2A_1 + A_0$.}
	\label{DACcir}
\end{figure*}

The DAC circuit includes a two-stage voltage divider circuit and an adder, see Fig. \ref{DACcir} for an illustration. The folding count $K$ mentioned earlier is formed by three output pins of the microprocessor, with $A_2$ forming the most significant bit of $K$ and $A_0$ forming the least significant bit. The minimum of $K$ is achieved when all three pins output a low level, i.e., $A_2A_1A_0=000$, which translates to a decimal value of $0$. The maximum value of $K$ is reached when all three pins output a high level, i.e., $A_2A_1A_0=111$, which translates to a decimal value of $7$. Therefore, the maximum folding count $K$ for the analog sampling hardware circuit is $7$. However, since the microprocessor cannot directly output decimal numbers and can only represent high and low levels using the binary values $1$ and $0$, we use a two-stage voltage divider circuit and an adder to design a DAC module. The DAC output is an integer voltage signal within the range of $0$ to $7$.

As shown in Fig. \ref{DACcir}, the two-stage voltage divider circuit includes three three-level voltage divider circuits. Each output pin for the folding count has a dedicated three-level voltage divider circuit. All three-level voltage dividers function identically, so when any of the three folding count output pins outputs a high level (i.e., $3.3$ V), the output after the three-level voltage divider circuit is $1$ V. The adder is built using an operational amplifier with an external gain circuit. Based on the binary-to-decimal conversion relationship, we construct the adder circuit with the highest bit A2 having a gain of $4$, the middle bit A1 having a gain of $2$, and the lowest bit A0 having a gain of $1$. The output voltage of the adder circuit corresponds to the decimal voltage value of the folding count $K$.

\subsection{Multiplexer}
\begin{figure}[!ht]
	\centering
	\includegraphics[width=0.48\textwidth]{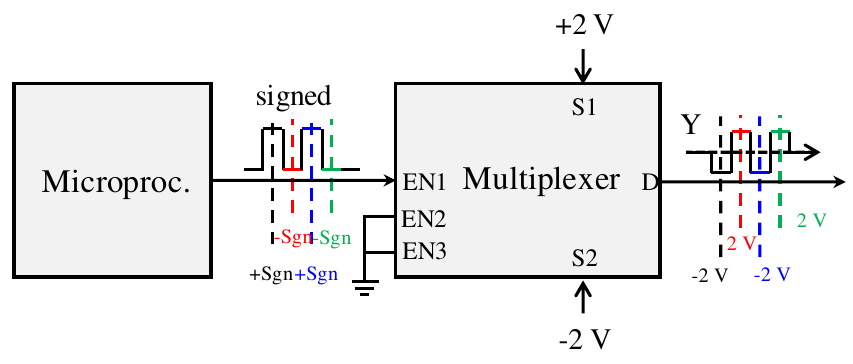}
	\caption{The Multiplexer circuit.}
	\label{Multiplexer}
\end{figure}

The multiplexer component functions as a multi-channel switch, allowing the selection of a corresponding channel’s output voltage by setting the high and low levels of a particular channel, see Fig. \ref{Multiplexer}. In the analog sampling hardware circuit, there are up to $8$ output channels provided by the multiplexer components. The channels are selected based on the high and low levels of the three control pins, EN1, EN2, and EN3. In this circuit, the threshold is set at $1$, meaning the single folding amount is $2$ V. The output pins are configured as follows: S1 is set to $+2$ V and S2 is set to $-2$ V. Both EN3 and EN2 pins are connected to low levels, and the selection of S1 or S2 is controlled by the EN1 pin. The EN1 pin is connected to the microprocessor's sign bit (Sgn). When Sgn outputs a high level (EN1 = $1$), the multiplexer selects the S2 pin, and the output pin D of the multiplexer outputs the single folding amount $Y=-2$ V. Conversely, when Sgn outputs a low level (EN1 $= 0$), the multiplexer selects the S1 pin, and the output pin D of the multiplexer outputs the single folding amount $Y=+2$ V.
\subsection{Multiplier}

\begin{figure}[htb!]
	\centering
	\includegraphics[width=0.48\textwidth]{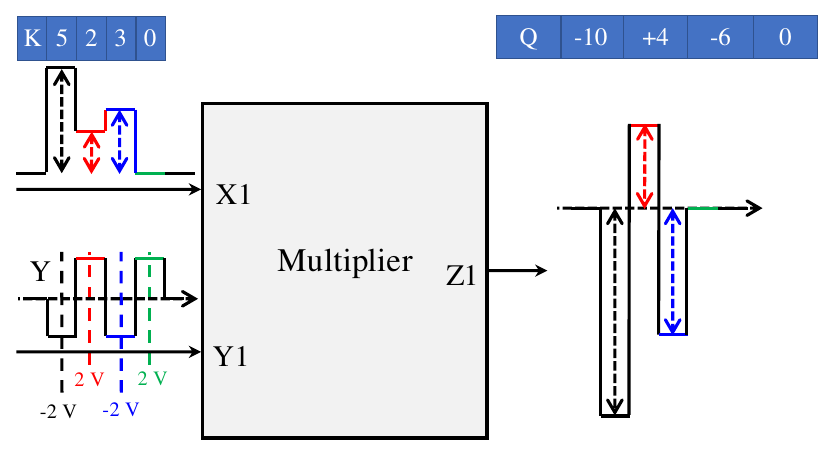}
	\caption{The Multiplier circuit.}
	\label{Multiplier}
\end{figure}

As shown in Fig. \ref{Multiplier}, the multiplier is primarily responsible for calculating the product of the folding count 
$K$ and the single folding amount $Y$. The folding count $K$ and the single folding amount 
$Y$ are input to the two inputs X1 and Y1 of the multiplier, respectively. The output Z1 of the multiplier provides the product of these two inputs, which represents the folding amount $Q$ of the input signal at a given moment. The implementation of the multiplier is relatively straightforward and will not be elaborated further in this text.

When the folding amount $Q$ of the input signal at a given moment is calculated, it needs to be added to or subtracted from the external input signal. This operation is performed by an adder. One input of the adder is the external input signal, and the other input is the folding amount $Q$. The result of this addition is the signal after one folding operation.
\subsection{Sampling the Folded Signal}\label{SecSampling}
\begin{figure*}[htb!]
	\centering
	\includegraphics[width=0.6\textwidth]{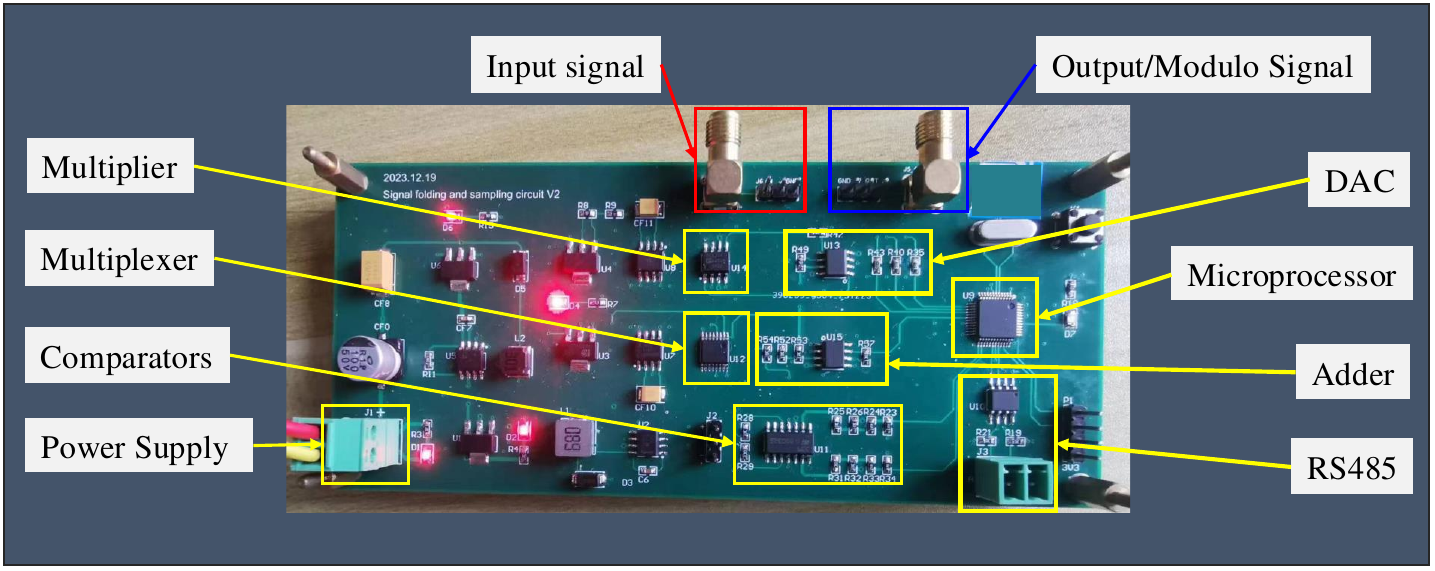}
	\caption{The hardware board for modulo ADC.}
	\label{HardwarePro}
\end{figure*}
\begin{figure}[htb!]
	\centering
\includegraphics[width=0.6\textwidth]{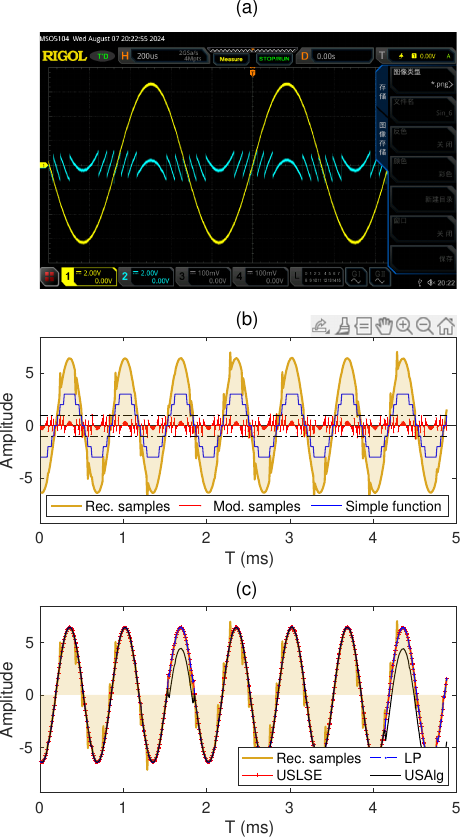}
	\caption{Results of $1.5$ kHz sine wave of amplitude $6.5$ V, and the maximum folding time is $3$. The oversampling factor is $\approx 34$. (a) Oscilloscope screenshots capturing the input signal and the folded signal. (b) Modulo samples, folding times for each sample, and the reconstructed samples from the modulo ADC. (c) Reconstruction results of the algorithms.}
    \label{Simu_Sin1}
\end{figure}
\begin{figure}[htb!]
	\centering
	\includegraphics[width=0.6\textwidth]{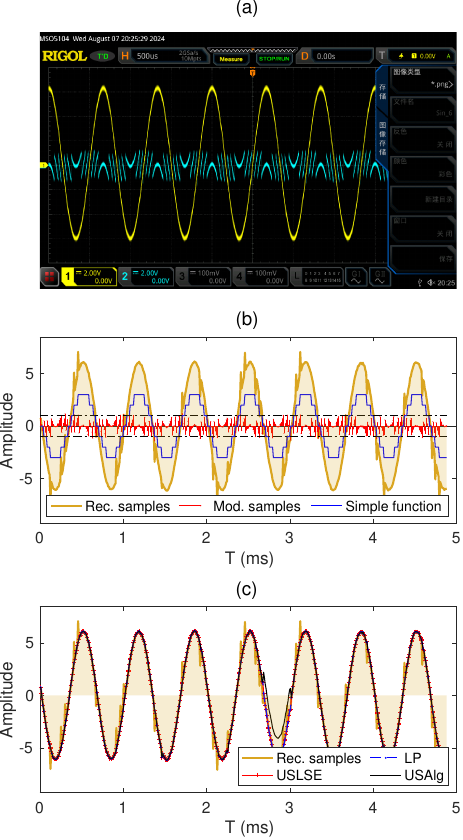}
	\caption{Results of $1.5$ kHz sine wave of amplitude $6.5$ V with additional $5\%$ noise, and the maximum folding time is $3$. The oversampling factor is $\approx 34$.}
    \label{Simu_Sin2}
\end{figure}
\begin{figure}[htb!]
	\centering
	\includegraphics[width=0.6\textwidth]{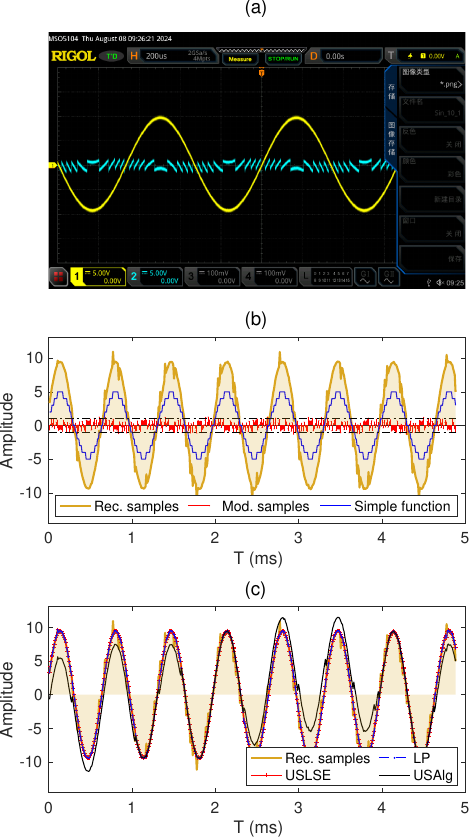}
	\caption{Results of $1.5$ kHz sine wave of amplitude $10$ V with additional $5\%$ noise, and the maximum folding time is $5$. The oversampling factor is $\approx 34$.}
    \label{Simu_Sin3}
\end{figure}
\begin{figure}[htb!]
	\centering
	\includegraphics[width=0.6\textwidth]{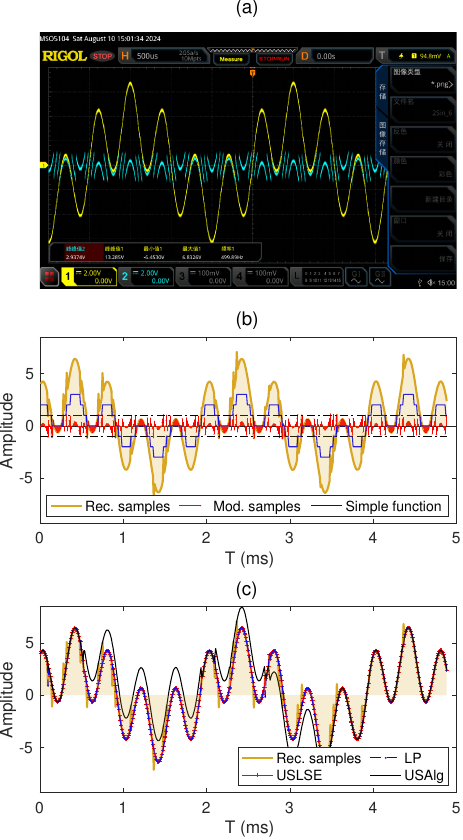}
	\caption{Results of superposition of two sine waves. We set $f_1=0.5$ kHz, $f_2 = 2.5$ kHz, and $A = 6.5$ V. The oversampling factor is $\approx 20$ and the maximum folding time is $3$.}
    \label{Simu_2Sin1}
\end{figure}
\begin{figure}[htb!]
	\centering
	\includegraphics[width=0.6\textwidth]{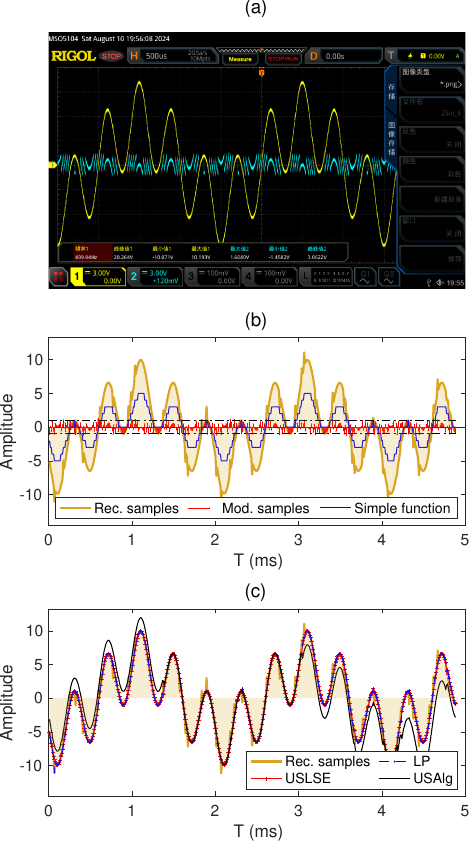}
	\caption{Results of superposition of two sine waves. We set $f_1=0.5$ kHz, $f_2 = 2.5$ kHz, and $A = 10$ V. The oversampling factor is $\approx 20$ and the maximum folding time is $5$.}
    \label{Simu_2Sin2}
\end{figure}
\begin{figure}[htb!]
	\centering
	\includegraphics[width=0.6\textwidth]{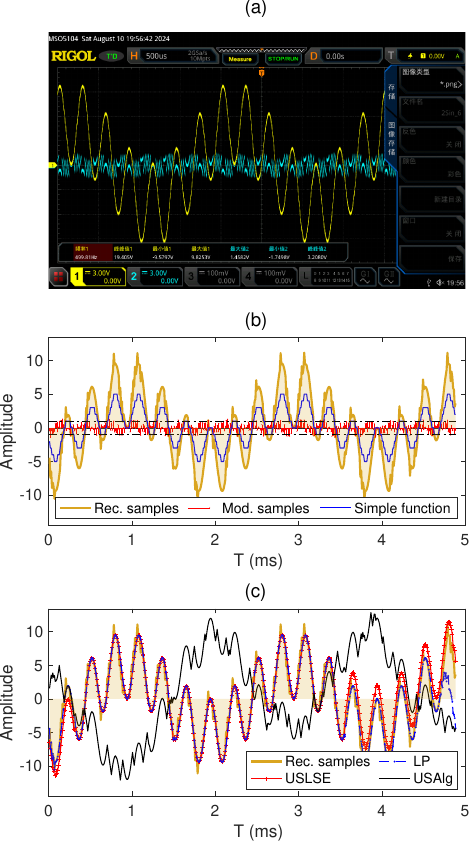}
	\caption{Results of superposition of two sine waves. We set $f_1=0.5$ kHz, $f_2 = 3.5$ kHz, and $A = 10$ V. The oversampling factor is $\approx 14$ and the maximum folding time is $5$.}
    \label{Simu_2Sin3}
\end{figure}
\begin{figure}[htb!]
	\centering
	\includegraphics[width=0.6\textwidth]{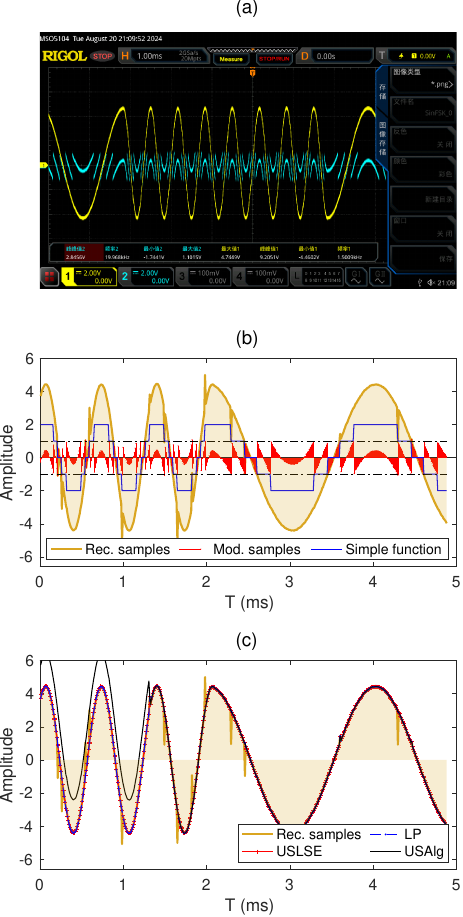}
	\caption{Results of FSK signal using a sine wave carrier with $f_1 = 0.5$ kHz, $f_2 = 1.5$ kHz, and $A = 4.5$ V. The oversampling factor is $\approx 34$ and the maximum folding time is $2$.}
    \label{Simu_SinFSK1}
\end{figure}
\begin{figure}[htb!]
	\centering
	\includegraphics[width=0.6\textwidth]{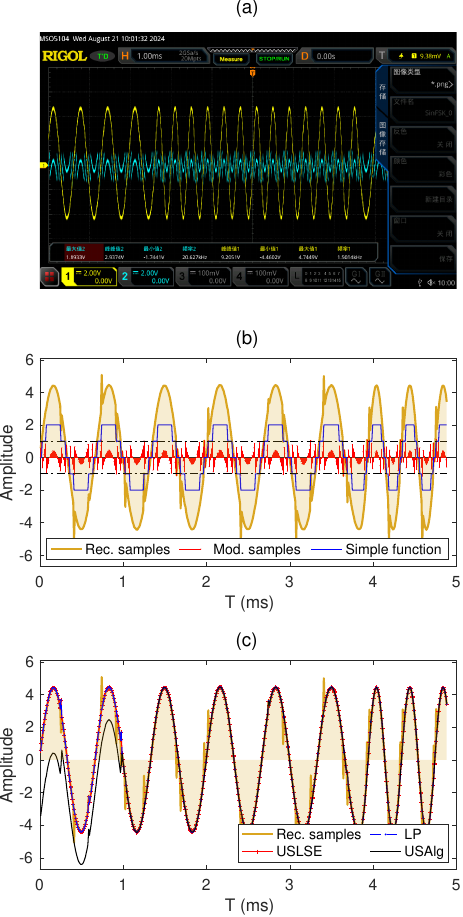}
	\caption{Results of FSK signal using a sine wave carrier with $f_1 = 1.5$ kHz, $f_2 = 2.5$ kHz, and $A = 4.5$ V. The oversampling factor is $\approx 20$ and the maximum folding time is $2$.}
    \label{Simu_SinFSK2}
\end{figure}
\begin{figure}[htb!]
	\centering
	\includegraphics[width=0.6\textwidth]{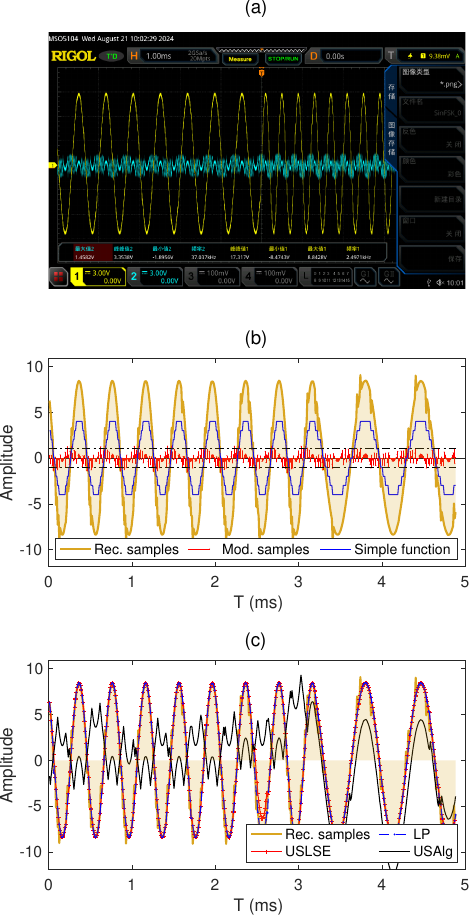}
	\caption{Results of FSK signal using a sine wave carrier with $f_1 = 1.5$ kHz, $f_2 = 2.5$ kHz, and $A = 8.5$ V. The oversampling factor is $\approx 20$ and the maximum folding time is $4$.}
    \label{Simu_SinFSK3}
\end{figure}
\begin{figure}[htb!]
	\centering
	\includegraphics[width=0.6\textwidth]{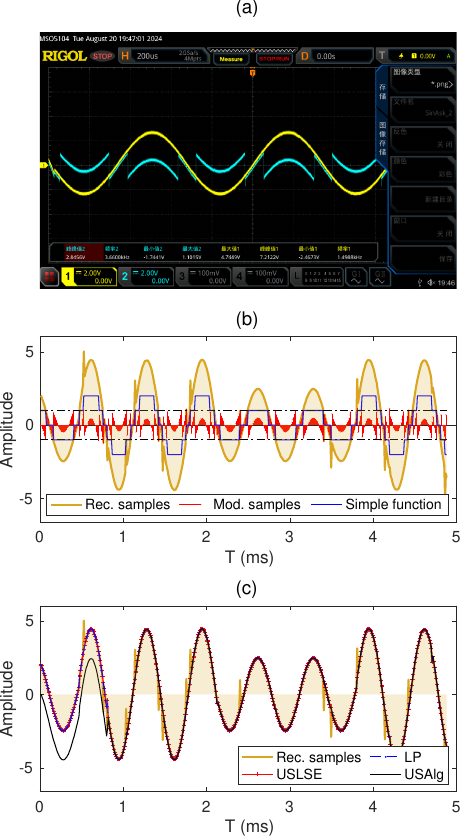}
	\caption{Results of ASK signal using a sine wave carrier with $f_c = 1.5$ kHz, $A_1 = 4.5$ V, and $A_2 = 2.5$ V. The oversampling factor is $\approx 34$ and the maximum folding time is $2$.}
    \label{Simu_SinASK1}
\end{figure}
\begin{figure}[htb!]
	\centering
	\includegraphics[width=0.6\textwidth]{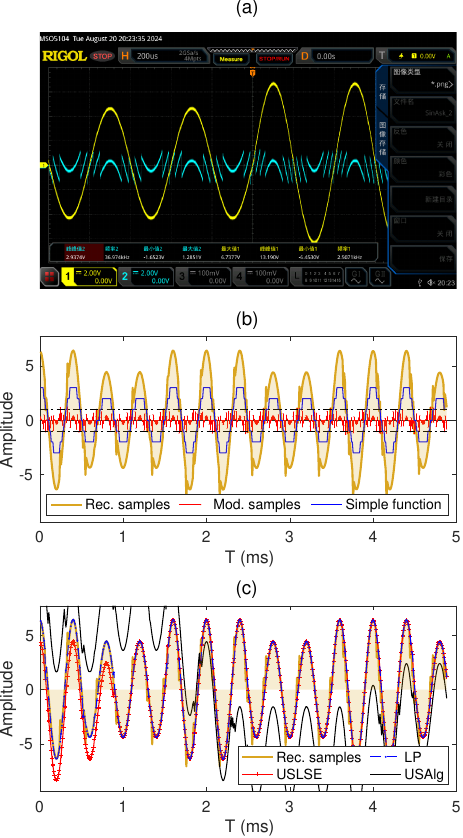}
	\caption{Results of ASK signal using a sine wave carrier with $f_c = 2.5$ kHz, $A_1 = 6.5$ V, and $A_2 = 4.5$ V. The oversampling factor is $\approx 20$ and the maximum folding time is $3$.}
    \label{Simu_SinASK2}
\end{figure}
\begin{figure}[htb!]
	\centering
	\includegraphics[width=0.6\textwidth]{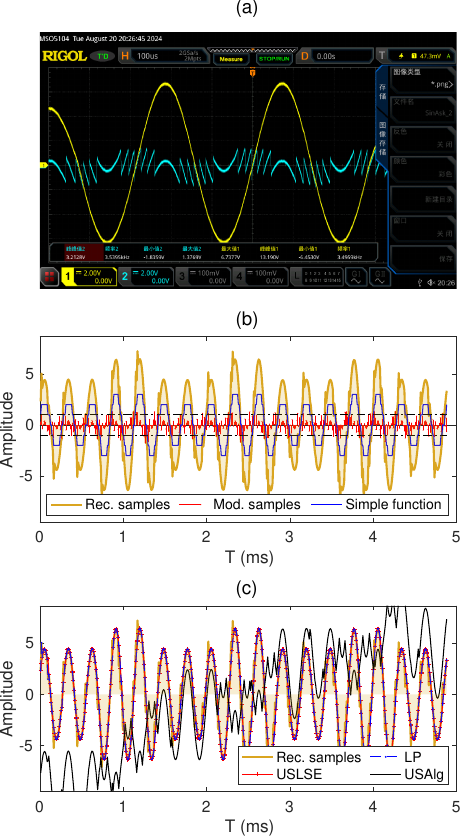}
	\caption{Results of ASK signal using a sine wave carrier with $f_c = 3.5$ kHz, $A_1 = 6.5$ V, and $A_2 = 4.5$ V. The oversampling factor is $\approx 14$ and the maximum folding time is $3$.}
    \label{Simu_SinASK3}
\end{figure}

The ADC is used to sample the folded signal. The built-in ADCs in the microprocessor capable of collecting negative voltage signals generally require bipolar power supplies. Obviously, for microprocessors with unipolar power supplies, their built-in ADCs cannot collect bipolar voltage signals. The microprocessor chip used in the analog sampling hardware circuit has a unipolar power supply with a supply voltage of $3.3$ V, so the ADC collection range is $0-3.3$ V. Therefore, a slight adjustment is needed before the ADC collects the folded signal. For folded signals within the range $[-1,1]$ V, as previously mentioned, the adder is used to add the folded signal to $+2$ V. This operation shifts the signal from the range $[-1,1]$ V to $[1,3]$ V. The microprocessor's built-in ADC then performs the signal sampling on this shifted signal when $S$ is positive. After sampling, we minus the bias $2$ V to obtain the modulo signal. It should be noted that although the ADC samples the modulo signal $x(t)$ within the range of $[-1, 1]$ V, the actual dynamic range of the ADC is $[-2, 1.3]$ V, due to its collection range of $0-3.3$ V.

In the hardware setup, the integrated ADC in the microprocessor has $12$-bit resolution, automatic calibration, customizable ADC conversion time, and other advantages. Additionally, we used timers and serial port interrupt functions to complement the microprocessor's ADC. The timer and serial port interrupt trigger the ADC collection, allowing experimenters to automatically adjust the sampling frequency, with a maximum sampling frequency of up to $12$ MHz. The ADC completes the collection of the folded signal, which is then transmitted to the host computer via the RS-$485$ bus. This facilitates the subsequent processing of the folded signal data by experimenters.

\subsection{Hardware Board for Modulo ADC}
Fig. \ref{HardwarePro} illustrates our modulo hardware board, highlighting the functions of its key components. The threshold of the modulo ADC is set as $\lambda = 1$ V. The details of the hardware components are listed in Table \ref{prototype}.
\begin{table}[htb!]
	\begin{center}
		\caption{The hardware components of the modulo sampling hardware prototype.}\label{prototype}
		\begin{tabular}{|c|c|c|}
			\hline
		Component& Model Number & Make  \\ \hline
			Comparator & LM339DR & Texas Instruments  \\ \hline
			Microprocessor & STM32F103C8T6 & STMicroelectronics \\ \hline
			Analog Mux & ADG1608BRUZ & Analog Devices  \\ \hline
			Analog Multiplier & AD623ARZ & Analog Devices  \\ \hline
			Adder& LT1364CN8\#PBF & Analog Devices  \\ \hline
		\end{tabular}
	\end{center}
\end{table}

\section{Real Experiments}
In this section, we evaluate the performance of our implemented modulo ADC. State-of-the-art algorithms, including our proposed USLSE \cite{zhang2024line}, LP \cite{ordentlich2018modulo}, and USAlg \cite{bhandari2020unlimited}, are employed to recover the original unfolded signal from modulo samples. Four types of signals are considered: a sine wave, a superposition of two sine waves, a frequency-shift keying (FSK) signal using a sine wave carrier, and an amplitude-shift keying (ASK) signal using a sine wave carrier. The threshold of the modulo ADC is set to $1$ V. As discussed in Sec. \ref{SecSampling}, the actual voltage range measurable by the ADC is $[-2, 1.3]$ V. The sampling frequency is $102.4$ kHz, and the sampling duration is $T = 4.87$ ms, resulting in a total of $500$ samples.
\subsection{Sine Wave}\label{SineWaveSimu}
In this section, we present the results for sine wave experiments. Fig. \ref{Simu_Sin1} illustrates the results for a $1.5$ kHz sine wave with an amplitude of $6.5$ V, resulting in an oversampling factor of approximately $34$ and the maximum folding times is $3$. Fig. \ref{Simu_Sin1}(a) displays the real-time capture from the oscilloscope, where the yellow line represents the original signal and the blue line represents the modulo signal. The asymmetry of the modulo signal is due to the time delay in the feedback loop of the modulo ADC, as discussed in Sec. \ref{HardProtoDe}. In Fig. \ref{Simu_Sin1}(b), the modulo samples $\mathbf{y}$ and the simple function $\boldsymbol{\epsilon}$ are shown. The reconstructed samples are directly calculated from the modulo samples and their respective folding instances. Some outliers are observed in the reconstructed samples, resulting from a timing mismatch between the modulo samples and folding instances due to hardware system delays. It should be noted that there are several modulo samples that exceed the range $[-1,1]$ V. Although the design ensures that the modulo signal $x(t)$ is sampled if and only if it lies within the range $[-1,1]$ V, in practice, due to non-ideal characteristics of the hardware, the actual sampled values do not strictly remain within this range. The reconstruction results of different algorithms are shown in Fig. \ref{Simu_Sin1}(c). USLSE and LP algorithms perfectly recover the original signal from the modulo samples while USAlg has some minor errors. Since the recovery relies solely on the modulo samples, the recovered signals of USLSE and LP do not exhibit the outliers seen in the reconstructed samples. 

We introduce an additional $5\%$ noise to the signal in Fig. \ref{Simu_Sin1}, and the result is presented in Fig. \ref{Simu_Sin2}. Fig. \ref{Simu_Sin2}(a) shows the real-time capture from the oscilloscope, where the modulo signal displays asymmetry due to the time delay in the feedback loop of the modulo ADC. Fig. \ref{Simu_Sin2}(b) illustrates the modulo samples, the folding times for each sample, and the reconstructed samples. The reconstruction results using different algorithms are shown in Fig. \ref{Simu_Sin2}(c). Both the USLSE and LP algorithms perfectly recover the original signal from the modulo samples. The USAlg algorithm almost perfectly recovered the signal, with some minor errors.

Next, we consider a sine wave with a greater amplitude. Fig. \ref{Simu_Sin3} presents the results for a $1.5$ kHz sine wave with an amplitude of $10$ V and additional $5\%$ noise, resulting in a maximum of $5$ folding times. Fig. \ref{Simu_Sin3}(a) shows the real-time capture from the oscilloscope, where the modulo signal exhibits asymmetry due to the time delay in the feedback loop of the modulo ADC. Fig. \ref{Simu_Sin3}(b) shows the modulo samples, folding times for each sample, and the reconstructed samples. The reconstruction results using different algorithms are shown in Fig. \ref{Simu_Sin3}(c). Both the USLSE and LP algorithms perfectly recover the original signal from the modulo samples, whereas USAlg fails.

\subsection{Superposition of Two Sine Waves}
In this section, we present the results for the superposition of two sine waves. Specifically, the original signal is given by $g(t) = \frac{A}{2}\sin(2\pi f_1 t) + \frac{A}{2}\sin(2\pi f_2 t)$, where $f_1$ is the frequency of the first sine wave and $f_2$ is the frequency of the second sine wave. The amplitudes of both sine waves are $A/2$.

Fig. \ref{Simu_2Sin1} illustrates the results for $f_1 = 0.5$ kHz, $f_2 = 2.5$ kHz, and $A = 6.5$ V. The oversampling factor is approximately $20$ and the maximum folding times is $3$. Fig. \ref{Simu_2Sin1}(a) displays the real-time capture from the oscilloscope. Compared to the sine wave signal in Sec. \ref{SineWaveSimu} with a frequency of $1.5$ kHz, the modulo signal exhibits severe asymmetry due to the increase in frequency of the original signal. In Fig. \ref{Simu_2Sin1}(b), the modulo samples $\mathbf{y}$, the simple function $\boldsymbol{\epsilon}$, and the reconstructed samples are shown. The reconstruction results of different algorithms are shown in Fig. \ref{Simu_2Sin1}(c). USLSE and LP algorithms perfectly recover the original signal from the modulo samples while USAlg fails. Now, we increase the amplitude $A$ from $6.5$ V to $10$ V, and the results are shown in Fig. \ref{Simu_2Sin2}. The oversampling factor is approximately $20$ and the maximum folding times is $5$. Fig. \ref{Simu_2Sin2}(a) shows the real-time capture from the oscilloscope, and Fig. \ref{Simu_2Sin2}(b) presents the modulo samples $\mathbf{y}$, the simple function $\boldsymbol{\epsilon}$, and the reconstructed samples.  Fig. \ref{Simu_2Sin2}(c) shows that USLSE and LP successfully recover the original samples from modulo samples while USAlg fails.

In Fig. \ref{Simu_2Sin3}, we present the result for the superposition of two sine waves with a higher frequency. We set $f_1 = 0.5$ kHz, $f_2 = 3.5$ kHz, and $A = 10$ V. The oversampling factor is approximately $14$ and the maximum folding times is $5$. Fig. \ref{Simu_2Sin3}(a) displays the real-time capture from the oscilloscope and Fig. \ref{Simu_2Sin3}(b) present the modulo samples, the simple function, and the reconstructed samples. It can be seen that the modulo signal displayed on the oscilloscope demonstrating significant asymmetry, and the dynamic range of the modulo signal is $[-1.75, 1.46]$ V which significantly exceeds the dynamic range of the modulo ADC, $[-1, 1]$ V. The dynamic range of the modulo samples is $[-1.45, 1.3]$ V. Due to certain non-ideal factors, although the modulo samples $\mathbf{y}$ do not strictly fall within the range of $[-1, 1]$ V, their dynamic range is narrower compared to the analog modulo signal $x(t)$. Fig. \ref{Simu_2Sin2}(c) shows that both USLSE and LP can almost recover the original signal, though with some errors, whereas USAlg struggles to recover the signal.

\subsection{FSK Signal Modulation}
In this section, we present the results for the FSK signal using a sine wave carrier, which is defined as 
\begin{align}
    g(t)= \begin{cases}A\sin \left(2 \pi f_1 t\right), & \text { if } b(t)=1 \\ A\sin \left(2 \pi f_2 t\right), & \text { if } b(t)=0\end{cases},
\end{align}
where $b(t)$ is the binary message signal, taking values $0$ or $1$, $f_1$ and $f_2$ are the frequencies associated with binary $1$ and binary $0$, respectively, and $A$ is the amplitude of the sine wave.

Fig. \ref{Simu_SinFSK1} illustrates the results for an FSK signal with $f_1 = 0.5 $ kHz, $ f_2 = 1.5 $ kHz, and $ A = 4.5 $ V. The oversampling factor is approximately $ 34 $, with a maximum of $ 2 $ folding times. In Fig. \ref{Simu_SinFSK1}(a), the oscilloscope's real-time capture is displayed. Fig. \ref{Simu_SinFSK1}(b) shows the modulo samples $ \mathbf{y} $, the simple function $ \boldsymbol{\epsilon} $, and the reconstructed samples. The reconstruction results are shown in Fig. \ref{Simu_SinFSK1}(c), where both USLSE and LP recover the original signal accurately, while USAlg demonstrates some minor errors. Next, we increase the carrier frequencies to $ f_1 = 1.5 $ kHz and $ f_2 = 2.5 $ kHz with $ A = 4.5 $ V, resulting in an oversampling factor of approximately $ 20 $. Fig. \ref{Simu_SinFSK2}(a) captures the oscilloscope output, and Fig. \ref{Simu_SinFSK2}(b) presents the modulo samples, the simple function, and the reconstructed samples. As shown in Fig. \ref{Simu_SinFSK2}(c), USLSE and LP effectively recover the original signal, while USAlg has some minor reconstruction errors. Finally, we increase the amplitude from $4.5$ V to $ A = 8.5 $ V resulting in a maximum folding times of $ 4 $, and the results are shown in Fig. \ref{Simu_SinFSK3}. Fig. \ref{Simu_SinFSK3}(a) displays the real-time oscilloscope capture, and Fig. \ref{Simu_SinFSK3}(b) shows the modulo samples, the simple function, and the reconstructed samples. In Fig. \ref{Simu_SinFSK3}(c), LP successfully recovers the original signal, USLSE nearly achieves full recovery, whereas USAlg struggles with signal reconstruction.

\subsection{ASK Signal Modulation}
In this section, we present the results for the ASK signal using a sine wave carrier, which is defined as 
\begin{align}
    g(t)= A(t) \sin \left(2 \pi f_c t\right),
\end{align}
where $A(t)$ which takes on values of either $A_1$ or $A_2$ and $f_c$ is the frequency of the sine wave.

Fig. \ref{Simu_SinASK1} presents the results for an ASK signal with a carrier frequency of $f_c = 1.5$ kHz and amplitudes $ A_1 = 4.5 $ V and $ A_2 = 2.5 $ V. The oversampling factor is approximately $34$, with a maximum of $2$ foldings. In Fig. \ref{Simu_SinASK1}(a), the real-time capture from the oscilloscope is displayed. Fig. \ref{Simu_SinASK1}(b) depicts the modulo samples, $\mathbf{y}$, the simple function $\boldsymbol{\epsilon}$, and the reconstructed samples. The reconstructed signal is illustrated in Fig. \ref{Simu_SinASK1}(c), where both USLSE and LP accurately recover the original signal, while USAlg shows minor discrepancies. Next, by increasing the carrier frequency to $f_c = 2.5$ kHz and adjusting the amplitudes to $ A_1 = 6.5 $ V and $ A_2 = 4.5 $ V, the oversampling factor decreases to approximately $20$, and the maximum number of foldings rises to $3$. These results are shown in Fig. \ref{Simu_SinASK2}. Fig. \ref{Simu_SinASK2}(a) captures the oscilloscope output, and Fig. \ref{Simu_SinASK2}(b) illustrates the modulo samples, the simple function, and the reconstructed samples. As seen in Fig. \ref{Simu_SinASK2}(c), LP effectively recovers the original signal, USLSE achieves near-complete recovery, but USAlg encounters difficulties in accurate reconstruction. Finally, increasing the carrier signal frequency in Fig. \ref{Simu_SinASK2} from $2.5$ kHz to $3.5$ kHz results in an oversampling factor of around $14$. The corresponding results are displayed in Fig. \ref{Simu_SinASK3}. Fig. \ref{Simu_SinASK3}(a) shows the real-time oscilloscope capture, and Fig. \ref{Simu_SinASK3}(b) provides the modulo samples, the simple function, and the reconstructed samples. As indicated in Fig. \ref{Simu_SinASK3}(c), both USLSE and LP effectively reconstruct the original signal, whereas USAlg experiences challenges in achieving accurate signal reconstruction.

\section{Conclusion}
In this paper, a modulo sampling hardware prototype is designed and several reconstruction algorithms are evaluated for reconstruction of different bandlimited signals. It is shown that the hardware is able to fold the signal, and the original signal is able to be recovered successfully when the dynamic range of the signal is $10$ times that of the ADC, and both LP and USLSE are robust in the presence of small noise. 

\bibliography{mybib}

\begin{thebibliography}{10}

\bibitem{OrdoñezfullduplexmoduloADC}
L.~G. Ordo{\~n}ez, P.~Ferrand, M.~Duarte, M.~Guillaud, and G.~Yang, ``On full-duplex radios with modulo-{ADC}s,'' {\em IEEE Open J. Commun. Soc.}, vol.~2, pp.~1279--1297, 2021.

\bibitem{MassiveMIMOlambda}
Z.~Liu, A.~Bhandari, and B.~Clerckx, ``$\lambda$–{MIMO}: {M}assive {MIMO} via modulo sampling,'' {\em IEEE Trans. Commun.}, vol.~71, no.~11, pp.~6301--6315, 2023.

\bibitem{sasagawa2015implantable}
K.~Sasagawa, T.~Yamaguchi, M.~Haruta, Y.~Sunaga, H.~Takehara, H.~Takehara, T.~Noda, T.~Tokuda, and J.~Ohta, ``An implantable {CMOS} image sensor with self-reset pixels for functional brain imaging,'' {\em IEEE Trans. Electron Devices}, vol.~63, no.~1, pp.~215--222, 2015.

\bibitem{ordentlich2018modulo}
O.~Ordentlich, G.~Tabak, P.~K. Hanumolu, A.~C. Singer, and G.~W. Wornell, ``A modulo-based architecture for analog-to-digital conversion,'' {\em IEEE J. Sel. Topics Signal Process.}, vol.~12, no.~5, pp.~825--840, 2018.

\bibitem{bhandari2020unlimited}
A.~Bhandari, F.~Krahmer, and R.~Raskar, ``On unlimited sampling and reconstruction,'' {\em IEEE Trans. Signal Process.}, vol.~69, pp.~3827--3839, 2020.

\bibitem{bhandari2019identifiability}
A.~Bhandari and F.~Krahmer, ``On identifiability in unlimited sampling,'' in {\em Proc. Int. Conf. Sampling Theory Appl.}, pp.~1--4, 2019.

\bibitem{romanov2019above}
E.~Romanov and O.~Ordentlich, ``Above the {N}yquist rate, modulo folding does not hurt,'' {\em IEEE Signal Process. Lett.}, vol.~26, no.~8, pp.~1167--1171, 2019.

\bibitem{prasanna2020identifiability}
D.~Prasanna, C.~Sriram, and C.~R. Murthy, ``On the identifiability of sparse vectors from modulo compressed sensing measurements,'' {\em IEEE Signal Process. Lett.}, vol.~28, pp.~131--134, 2020.

\bibitem{ordentlich2016integer}
O.~Ordentlich and U.~Erez, ``Integer-forcing source coding,'' {\em IEEE Trans. Inf. Theory}, vol.~63, no.~2, pp.~1253--1269, 2016.

\bibitem{APB2020}
E.~Domanovitz and U.~Erez, ``Achievability performance bounds for integer-forcing source coding,'' {\em IEEE Trans. Inf. Theory}, vol.~66, no.~3, pp.~1482--1496, 2020.

\bibitem{weiss2022blind}
A.~Weiss, E.~Huang, O.~Ordentlich, and G.~W. Wornell, ``Blind modulo analog-to-digital conversion,'' {\em IEEE Trans. Signal Process.}, vol.~70, pp.~4586--4601, 2022.

\bibitem{romanov2021blind}
E.~Romanov and O.~Ordentlich, ``Blind unwrapping of modulo reduced {G}aussian vectors: {R}ecovering {MSB}s from {LSB}s,'' {\em IEEE Trans. Inf. Theory}, vol.~67, no.~3, pp.~1897--1919, 2021.

\bibitem{mulleti2024modulo}
S.~Mulleti and Y.~C. Eldar, ``Modulo sampling of {FRI} signals,'' {\em IEEE Access}, vol.~12, pp.~60369--60384, 2024.

\bibitem{Qi2023ID}
Q.~Zhang, J.~Zhu, F.~Qu, Z.~Zhu, and D.~W. Soh, ``On the identifiability from modulo measurements under {DFT} sensing matrix,'' {\em arXiv preprint arXiv:2401.00194}, 2023.

\bibitem{cheng2023crb}
Y.~Cheng, J.~Karlsson, and J.~Li, ``Cram\'er-rao bound for signal parameter estimation from modulo {ADC} generated data,'' {\em IEEE Trans. on signal Process.}, 2024.

\bibitem{bhandari2018unlimited1}
A.~Bhandari, F.~Krahmer, and R.~Raskar, ``Unlimited sampling of sparse sinusoidal mixtures,'' in {\em Proc. IEEE Int. Symp. Info. Theory}, pp.~336--340, 2018.

\bibitem{bhandari2018unlimited2}
A.~Bhandari, F.~Krahmer, and R.~Raskar, ``Unlimited sampling of sparse signals,'' in {\em Proc. IEEE Int. Conf. Acoust., Speech Signal Process.}, pp.~4569--4573, 2018.

\bibitem{bouis2021multidimensional}
V.~Bouis, F.~Krahmer, and A.~Bhandari, ``Multidimensional unlimited sampling: {A} geometrical perspective,'' in {\em Proc. Eur. Signal Process. Conf.}, pp.~2314--2318, 2021.

\bibitem{fernandez2021doa}
S.~Fern{\'a}ndez-Menduina, F.~Krahmer, G.~Leus, and A.~Bhandari, ``Do{A} estimation via unlimited sensing,'' in {\em Proc. Eur. Signal Process. Conf.}, pp.~1866--1870, 2021.

\bibitem{bhandari2021modulo}
A.~Bhandari, M.~Beckmann, and F.~Krahmer, ``The modulo {R}adon transform and its inversion,'' in {\em Proc. Eur. Signal Process. Conf.}, pp.~770--774, 2021.

\bibitem{fernandez2022computational}
S.~Fern{\'a}ndez-Mendui{\~n}a, F.~Krahmer, G.~Leus, and A.~Bhandari, ``Computational array signal processing via modulo non-linearities,'' {\em IEEE Trans. Signal Process.}, vol.~70, pp.~2168--2179, 2022.

\bibitem{beckmann2020hdr}
M.~Beckmann, F.~Krahmer, and A.~Bhandari, ``{HDR} tomography via modulo {Radon} transform,'' in {\em Proc. IEEE Int. Conf. Image Process.}, pp.~3025--3029, 2020.

\bibitem{bhandari2020hdr}
A.~Bhandari and F.~Krahmer, ``{HDR} imaging from quantization noise,'' in {\em Proc. IEEE Int. Conf. Image Process.}, pp.~101--105, 2020.

\bibitem{zhang2024line}
Q.~Zhang, J.~Zhu, F.~Qu, and D.~W. Soh, ``Line spectral estimation via unlimited sampling,'' {\em IEEE Trans. Aerosp. Electron. Syst.}, 2024.

\bibitem{Qi2023onebit}
Q.~Zhang, J.~Zhu, F.~Qu, and D.~W. Soh, ``One-bit-aided modulo sampling for {DOA} estimation,'' {\em arXiv preprint arXiv:2309.04901}, 2023.

\bibitem{bhandari2021unlimited}
A.~Bhandari, F.~Krahmer, and T.~Poskitt, ``Unlimited sampling from theory to practice: {F}ourier-{P}rony recovery and prototype {ADC},'' {\em IEEE Trans. Signal Process.}, vol.~70, pp.~1131--1141, 2021.

\bibitem{azar2022residual}
E.~Azar, S.~Mulleti, and Y.~C. Eldar, ``Residual recovery algorithm for modulo sampling,'' in {\em Proc. IEEE Int. Conf. Acoust., Speech Signal Process.}, pp.~5722--5726, 2022.

\bibitem{azar2022robust}
E.~Azar, S.~Mulleti, and Y.~C. Eldar, ``Robust unlimited sampling beyond modulo,'' {\em arXiv preprint arXiv :2206.14656}, 2022.

\bibitem{florescu2022surprising}
D.~Florescu, F.~Krahmer, and A.~Bhandari, ``The surprising benefits of hysteresis in unlimited sampling: {T}heory, algorithms and experiments,'' {\em IEEE Trans. Signal Process.}, vol.~70, pp.~616--630, 2022.

\bibitem{florescu2022time}
D.~Florescu and A.~Bhandari, ``Time encoding via unlimited sampling: {T}heory, algorithms and hardware validation,'' {\em IEEE Trans. on Signal Process.}, vol.~70, pp.~4912--4924, 2022.

\bibitem{Krishnacircuit2019}
A.~Krishna, S.~Rudresh, V.~Shaw, H.~R. Sabbella, C.~S. Seelamantula, and C.~S. Thakur, ``Unlimited dynamic range analog-to-digital conversion,'' {\em arXiv preprint arXiv:1911.09371}, 2019.

\bibitem{mulleti2023hardware}
S.~Mulleti, E.~Reznitskiy, S.~Savariego, M.~Namer, N.~Glazer, and Y.~C. Eldar, ``A hardware prototype of wideband high-dynamic range analog-to-digital converter,'' {\em IET Circuits, Devices Syst.}, vol.~17, no.~4, pp.~181--192, 2023.

\end{thebibliography}
\bibliographystyle{ieeetr}
\end{document}